\documentclass[12pt]{iopart}

\usepackage{iopams}

\usepackage[T2A]{fontenc}
\usepackage[english]{babel}

\def\bi{\bibitem}
\def\rN{{\rm N}}

\def\rs{{\rm s}}       \def\rc{{\rm c}}      
\def\rcs{{\rm cs}}

\def\nc{{\rm nc}}
\def\bs{\boldsymbol}

\def\ome{\omega}

\def\alf{\alpha}

\def\Dlt{\Delta}
\def\dlt{\delta}

\def\lam{\lambda}
\def\vphi{\varphi}
\def\sig{\sigma}
\def\eps{\varepsilon}
\def\dg{\dagger}
\def\ra{\rangle}
\def\la{\langle}
\def\fns{\footnotesize}
\def\psl{\partial}

\def\rt{\right}
\def\dstyle{\displaystyle}
\def\ovl{\overline}   \def\tld{\tilde} 
\def\lb{\label}
\def\beq{\begin{equation}}
\def\eeq{\end{equation}}
\def\bea{\begin{eqnarray}}
\def\eea{\end{eqnarray}}
\def\bt{\begin{tabular}}
\def\et{\end{tabular}}
\def\ftm{\phantom{}}
\newcommand{\rd}{\mathrm{d}}            
\newcommand{\ri}{\mathrm{i}}

\begin{document}

\title[Uncertainty Relations in Stochastic Mechanics]
{On the Uncertainty Relations in Stochastic Mechanics}
\author{D.A.  Trifonov\,$^1$, B.A. Nikolov\,$^2$,  \MakeLowercase{and}
I.M. Mladenov $^3$}
\address{$^1$ Institute of Nuclear Research, Bulgarian Academy of Sciences,\\
 Tzarigradsko chaussee 72, 1184 Sofia, Bulgaria}
\address{$^2$ Department of Physics, Konstantin Preslavski University, \\
9712 Shumen, Bulgaria}
\address{$^3$ Institute of Biophysics, Bulgarian Academy of Sciences,\\
Acad. G. Bonchev Str., Bl. 21, 1113 Sofia, Bulgaria}

\vspace{-6.5cm}
\noindent e-print quant-ph/0902.3880v3\\
\vspace{6.5cm}

\begin{abstract} \footnote[7]{Amended version of the withdrawn e-print quant-ph/0902.3880v1}

It is shown that the Bohm equations for the phase $S$ and squared modulus $\rho$ 
of the quantum mechanical wave function can be derived from the classical ensemble 
equations admiting an aditional momentum $p_s$ of the form proportional to the osmotic 
velocity in the Nelson stochastic mechanics and using the variational principle with 
appropriate change of variables. The possibility to treat  grad$S$ and $p_s$ as 
two parts of the momentum of quantum ensemble particles is considered from the 
view point of uncertainty relations of Robertson - Schroedinger type on the examples 
of the stochastic image of quantum mechanical canonical coherent and squeezed states. 
\end{abstract}


\section{Introduction} 

The uncertainty (indeterminacy) principle in quantum physics, which quantitatively is 
expressed in the form of  uncertainty relations (URs) \cite{H} - \cite{S} is commonly 
regarded as the most radical departure from the classical physics.

However in the recent decades publications have appeared %
\cite {Cetto72} - \cite{Hall02} in which inequalities are introduced in Nelson stochastic  
mechanics (SM) \cite{Nelson66} and discussed as Heisenberg-type URs .
The equations of motion of this mechanics coincide with the David Bohm equations 
\cite{Bohm52} (the continuity equation  and the modified  Hamilton-Jacobi equation, 
the latter known also as  Hamilton-Jacobi-Madelung (HJM) equation)) for the phase 
$S$ and squared modulus $|\psi|^2\equiv \rho$ of the Schr\"odinger wave function $\psi$. 
Bohm equations for  $ S$ and $\rho$ have been later derived from `the stochastic 
variational principles of control theory'  by  Guerra and Marra \cite{Guerra83}, and 
by Reginatto \cite{Regi98}, using the `principle of minimum Fisher information' \cite{Frieden}.

 Hall and Reginatto \cite{Hall02} introduced the so called `exact  UR' and showed 
that it `leads from classical equations of motion to the Schr\"odinger equation' and to  
uncertainty inequality of the form of Heisenberg UR. 
They derived the continuity equation and the modified HJ equation  from a variational principle, 
 introducing into the Lagrangian of the HJ equation an additional   momentum $p_{\rN}$ 
 of the classical particle assuming that its first moment and its covariance with the `classical' 
moment $\nabla S$ are universally vanishing. 
From some general consideration they `derived' that the variance of this extra momentum 
should be proportional to the Fisher information of the coordinate probability density $\rho$. 
As a result their Lagrangian  takes the form of Reginatto' Lagrangian, wherefrom the 
Bohm equations are derived  \cite{Regi98} and the product of coordinate and 
$\bs p_{\rN}$ variances equals constant for any $\rho$ (which in fact is a minimization 
of the Cramer-Rao inequality). 
This equality  and the related uncertainty principle are called `exact' UR and `exact' 
uncertainty principle. 
The system described by the so derived Bohm equations is interpreted in \cite{Hall02} 
as `{\it quantum ensemble}'. 
If $\nabla S/m$ and the variance of $\bs p_{\rN}/m$ are  identified with the current %
velocity and the mean squared osmotic velocity  the formal connection to the Nelson SM %
is established \cite{Hall02}.   
 
However no particular underlying   physical model was assumed for the fluctuations of 
the momentum $\bs p_\rN$ - they were regarded as fundamentally nonanalyzable \cite{Hall02}. 
Having no model for $\bs p_\rN$ one has to postulate infinitely many constraints in order to 
recover the statistical properties of quantum mechanical  momentum $\hat{p}$. %
The two constraints postulated in \cite{Hall02} (namely $\overline{\bs p_{\rN}} = 0$ and 
$\overline{\nabla S\cdot \bs p_{\rN}} = 0$) ensure the coinsidence only of the first %
two moments of $\nabla S + \bs p_{\rN}$ and $\hat{p}$. 

It is our aim here to introduce a model for such additional momentum to account partially 
for the quantum fluctuations and to examin its properties and consistency. 
(`Partially', because no classical model, we believe, could provide the `full' account).
Another our aim  is to briefly review the URs in the Nelson stochastic mechanics (SM) 
 from the point of view of the more precise Robertson-Schr\"odinger (R-S) inequality. 
Unlike the Heisenberg UR the R-S UR in quantum mechanics involves all the three second
moments of the two quantum observables {\small $\hat{A}$} and {\small $\hat{B}$}, the variances 
{\small$(\Dlt\hat A)^2$},  {\small $(\Dlt\hat B)^2$} and the squared covariance $(\Dlt_{AB})^2$. 
If the covariance is vanishing then the R-S UR recovers the Heisenberg UR.

In the next section we briefly review the Heisenberg and Robertson-Schr\"odinger URs (R-S UR). 
In the third section we recall the main features of the Hall and Reginatto `quantum ensemble' %
approach and  Nelson SM. 
A model of additional momentum $p_\rs$ is introduced, its potential $S_\rs$ being 
interpreted as the intensity dependent part of the quantum wave function phase $S$.
It is shown that Bohm equations for $S$ and $ \rho$ can be derived from the Reginatto 
variational principle considering the probability density $\rho$ and $S_- \equiv S-S_\rs$ 
as new independent variables.
In section 4 the stochastic  analogues of the R-S URs are reviewed and discussed 
in connection with the introduced auxilliary momentum model $p_s$ and on the example
of $S$ and $\rho$ corresponding to canonical coherent and squeezed states. 
In section 5   the first and second moments of coordinate and the related momenta and URs 
are calculated on the example of stochastic images of  canonical coherent and  squeezed 
states (CS and SS) and compared with the corresponding moments and URs in quantum %
mechanics. 
Nelson SM images of  canonical CS and SS have been discussed previously 
in several  papers: of CS in \cite{Guerra81} - \cite{Martino95}, %
and of  SS and CS - in \cite{Petroni99} - \cite{Martino95} in the context of  
`stochastic mechanics and control theory'.

\section{Robertson-Schr\"odinger UR in Quantum Mechanics}  

The indeterminacy principle was introduced in 1927 by Heisenberg  \cite{H} who 
demonstrated the impossibility of simultaneous precise  measurement of the canonical 
quantum observables $\hat{x}$ and $\hat{p}$ (the particles coordinate and momentum) %
by positing an approximate relation $\dlt p \,\dlt x \sim \hbar$,  where $\hbar$ is the Plank constant. 

Heisenberg considered this inequality as the ``direct descriptive interpretation'' of the
canonical commutation relation between the operators of the coordinate and momentum: 
$[\hat{x},\hat{p}] = \ri\hbar$, $[\hat{x},\hat{p}] \equiv \hat{x}\hat{p}-\hat{p} \hat{x}$. 
A rigorous proof of the Heisenberg relation was soon published by Kennard and Weyl 
\cite{KW} who established the inequality
\beq\lb{HUR} 
(\Dlt \hat p)^2 (\Dlt \hat x)^2 \,\geq\, \hbar/4, 
\eeq 
where $(\Dlt \hat p)^2$ and $(\Dlt \hat x)^2$ are the variances (dispersions) of $\hat{p}$ and
$\hat{x}$, defined by Weyl for every quantum state $|\psi\ra$ via the formula 
{\small $(\Dlt \hat p)^2:=  \la\psi|(\hat{p}-\la\psi|\hat{p}|\psi\ra)^2|\psi\ra,$} and
similarly  $(\Dlt\hat{x})^2$ is defined. In correspondence with the classical probability 
theory the standard deviation {\small $\Dlt \hat A$}  is considered as a measure for the 
uncertainty (indeterminacy) of the quantum observable {\small $\hat{A}$}  in the 
corresponding state $|\psi\ra$. 
The inequality (\ref{HUR}) became known as the {\it Heisenberg} UR.

The extension of (\ref{HUR}) to the case of two arbitrary quantum observables 
(Hermitian operators {\small $\hat{A}$} and {\small $\hat{B}$}) was made by Robertson and 
Schr\"odinger \cite{R, S}, who established more precise inequality, that involves all the 
three second moments of the two observables,
\beq\lb{SUR}
(\Dlt \hat A)^2(\Dlt \hat B)^2 - (\Dlt_{AB})^2 \,\,\geq\,\,
\frac{1}{4}\left|\la [\hat{A},\hat{B}]\ra\right|^2,
 \eeq where
$\Dlt_{AB}$ is the covariance (in mathematical  literature denoted
usually as Cov{\small$(AB)$} ) of {\small$\hat{A}$} and {\small$\hat{B}$},\,\,\,
{\small$\Dlt_{AB} := (1/2)\la \hat{A}\hat{B}+\hat{B}\hat{A}\ra - \la
\hat{A}\ra\la\hat{B}\ra$}.

In the case of coordinate and momentum observables relation (\ref{SUR}) takes the shorter  
form 
\beq \label{xp SUR} (\Dlt\hat x)^2(\Dlt \hat p)^2 - (\Dlt_{xp})^2 \,\,\geq\,\, \hbar^2/4. 
\eeq

The  inequality (\ref{SUR}) is referred to as Schr\"odinger or Robertson-Schr\"odinger UR  
(R-S UR). 
In states with vanishing covariance the R-S UR (\ref{xp SUR}) recovers the Heisenberg's one,
eq. (\ref{HUR}).  The minimization of (\ref{HUR}), i.e. the equality in (\ref{HUR}), means the 
equality in (\ref{xp SUR}), the inverse being not true. 
Thus the R-S UR provides a more stringent limitation (from below) to the product %
of two variances.  Besides the R-S UR is more symmetric than the Heisenberg UR: %
the equality in it is invariant under nondegenerate linear transformations of the %
two observables (in the case of $x$ and $p$ R-S UR is invariant under linear canonical %
transformations) \cite{T01}.
Despite these  advantages the relation (\ref{xp SUR}) and/or (\ref{SUR}) are lacking %
in almost all quantum mechanics text books.  %
The interest in R-S relation has been renewed in the last three %
decades  \cite{DKM, T94, Dod02} (50 years after its discovery) in connection with %
the description and experimental realization of the {\it squeezed states} of the %
electromagnetic radiation (see the `squeezed review' \cite{T01, Dod02}).

\section{`Quantum Ensemble' and Stochastic Mechanics}  

The quantum-classical relations are subject of a host of publications, which started from the early 
days of quantum mechanics. Aiming to provide an alternative  interpretation of quantum mechanics 
in terms of `hidden variables' David Bohm \cite{Bohm52} noted that the phase $S= \hbar \arg\psi$ 
and the squared modulus $|\psi|^2\equiv \rho$ of  the quantum-mechanical particle wave function $\psi$  
obeys a system of classical-type equations, namely the probability conservation equation 
and a modified Hamilton-Jacobi  equation,
\beq\label{Bohm eqs}
\frac{\psl \rho}{\psl t} + \frac{1}{m}{\rm div} \left(\rho\nabla S \right)  = 0, \qquad 
\frac{\psl S}{\psl t} + \frac{1}{2m}\left(\nabla S \right)^2 + V({\bs x},t) + V_{\rm q}
= 0,
\eeq
where $V$ is the external particle potential, and $V_{\rm q}$ (called  
`quantum potential'  \cite{Bohm52}) is given by
$$V_{\rm q} = \frac{\hbar^2}{8m}\left(\frac{(\nabla\rho)^2}{\rho^2} - 2\frac{\nabla^2\rho}{\rho}\right).$$
 Pursuing the classical interpretation and derivation of the Schr\"odinger equation Nelson \cite{Nelson66} 
derived equations for the velocity fields in the forward and backward Fokker-Planck equations of a 
diffusion process and, noting that the `osmotic' velocity $\bs u$ is a gradient 
($\bs u = D\nabla \ln\rho$, $\rho$ being the probability density of the process, $D$ - the diffusion %
coefficient) and supposing that the current velocity ${\bs v}$  is also a gradient, 
${\bs v} = (1/m) \nabla S$,  he had established that with $D = \hbar/2m$ the probability density %
$\rho$ and the current velocity potential $S$ satisfy the Bohm equations (\ref{Bohm eqs}),  %
i.e. $\psi := \sqrt{\rho}\exp(\ri S/\hbar)$ obey the Schr\"odinger equation. 
This theory is known as Nelson SM.  

Reginatto \cite{Regi98} noted that the Bohm equations (and thereby the Schr\"odinger equation) 
can be obtained from the variational principle and the principle of minimum Fisher information 
\cite{Frieden} applied to the `classical ensemble of particles'.  
In this derivation Reginatto started from the classical Hamilton-Jacobi (HJ) equation (we consider 
the case of $n=1$, and external potential $V$) 
\beq\label{HJ eq} \frac{\psl S}{\psl t} + \frac{1}{2m} \left(
\nabla S \right)^2  +  V({\bs x},t)   = 0. 
\eeq 
Supposing that the coordinates are subject to fluctuations described by the probability density 
$\rho$ he had postulated the validity of the continuity equation of the same form as in (\ref{Bohm eqs}),
\beq\lb{ceq1}
 \frac{\psl \rho}{\psl t}+ \frac{1}{m}{\rm div}
 \left(\rho\nabla S \right)  = 0
 \eeq
and noted that it can be derived from the functional
\beq\label{Phi_A}
 \Phi_A = \int \rho\,\left(\psl_t S +
\frac{1}{2m}\nabla S\cdot \nabla S + V\right) \rd^3 x\, \rd t 
\eeq 
as extremal equation with respect to variation of the classical action $S$. %
(As noted in \cite{Regi98} the variation with respect to $\rho$ trivially results %
into HJ eq. (\ref{HJ eq})). Physical system  which motion is described by the %
equations (\ref{HJ eq}) and (\ref{ceq1}) is called  {\it classical ensemble} %
of particles \cite{Regi98, Hall02}. 

To obtain the second of the Bohm equations (\ref{Bohm eqs}) the principle of %
minimal Fisher information was applied by adding to $\Phi_A$ the term \cite{Regi98} 
\beq\label{Phi_A'} 
\Phi_A^\prime =
\lam \int I_{\mbox{\tiny F}}(\rho) \rd t ,\qquad   I_{\mbox{\tiny F}} = 
\int \frac{1}{\rho} \nabla \rho\cdot \nabla \rho \, \rd^3x 
 \eeq
where $I_{\mbox{\tiny F}}$ is the Fisher information of the probability %
density $\rho({\bs x},t)$, and the multiplier $\lam$ is put equal to $\hbar^2/8m$. %
Thus the Bohm equations (\ref{Bohm eqs}) are derived from the action functional 
\bea\label{Phi_B} 
\Phi_B &= \Phi_A + \Phi^\prime_A \nonumber \\
           &=  \int \left[\rho\,\left(\psl_t S + \frac{1}{2m}\nabla S \cdot \nabla S +
V \right)  + \frac{\hbar^2}{8m}   \frac{1}{\rho} \nabla \rho\cdot \nabla \rho  \right]\, 
\rd^3 x\rd t 
\eea
by independent variation of $\rho$ and $S$. In fact  Bohm equations  have been derived %
previously from the same action functional (\ref{Phi_B}) by Guerra and Moratto  %
\cite{Guerra83} but with no reference to Fisher information.  

By different argumentation the same action functional (\ref{Phi_B}) has been derived and %
used later in \cite{Hall02}, where   the term $2m\lam I_{\mbox{\tiny F}}$ is interpreting %
 as a variance $(\sig_{p_{\rN}})^2$ of an additional momentum ${\bs p_{\rN}}$, subjected %
to the constraints  (of vanishing first moment and vanishing covariance with $\nabla S$)
\beq\label{N} 
\la {\bs p_{\rN}} \ra = 0,\qquad  \la {\bs p_{\rN}}\cdot \nabla S\ra = 0. 
\eeq
This variance obeys the inequality  (in the one-dimensional case)
\beq\label{ineq0} 
 (\sig_x)^2 (\sig_{p_{\rN}})^2 \geq  \hbar^2/4,
\eeq 
which directly stems from well known Cramer-Rao inequality 
$ (\sig_x)^2 I_{\mbox{\tiny F}} \geq 1$, 
where $ (\sig_x)^2$ is the variance (the squared uncertainty) of $x$.  The authors of  
\cite{Hall02} consider the total momentum of the particle $p$ as a sum of
$\psl_x S$ and  $p_{\rN}$. Then, in view of (\ref{N}) and (\ref{ineq0}), one gets 
\beq\label{HUR0}
 (\sig_x)^2 (\sig_p)^2 \geq  (\sig_x)^2 (\sig_{p_{\rN}})^2 \geq  \hbar^2/4 .
\eeq 
The authors argue that this is a derivation of Heisenberg UR.

However no particular underlying physical model was assumed for the  fluctuations of %
the momentum ${\bs p_{\rN}}$ - they were regarded as `fundamentally nonanalyzable'.  %
Despite the proclaimed `nonanalizability'  of  ${\bs p_{\rN}}$  the authors of %
 \cite{Hall02} succeeded to find that  its variance should be proportional to the %
Fisher information of $\rho$, resorting in this way to the Reginatto functional %
(\ref{Phi_B}), wherefrom they derive the Bohm equations for $\rho$ and $S$.    %
The so derived  equations (\ref{Bohm eqs}) are referred to as equations of motion %
of {\it quantum  ensemble} \cite{Regi98, Hall02}.

We note that in fact the second constraint in (\ref{N})  simply reduces the functional %
(6) in \cite{Hall02} to Reginatto functional (\ref{Phi_B}), which ensures the %
variational derivation of the Bohm equations for $\rho$ and $S$. %
Retaining the idea of introducing an additional stochastic momentum however there %
is an alternative way to derive variationally the Bohm equations, namely the change %
the independent variables. %
In this way one may expect to introduce a model of  an {\it analizable} additional %
stochastic momentum $p_{\rs}$ to account {\it partially} for the quantum momentum %
fluctuations. 

We consider the total momentum $\bs{p}$ of the particle as a sum of two parts, 
\beq\label{p_t} 
\bs{p} = \bs{p}_\rc + \bs{p}_\rs, 
\eeq 
supposing that the first one stems from the deterministic classical motion and the %
second one is induced by the coordinate randomization. We suppose further that both %
$\bs{p}_\rc$ and $\bs{p}_\rs$ are gradients of corresponding potentials (actions)
\beq\label{p_c, p_o} 
\bs{p}_\rc = \nabla S, \qquad  \bs{p}_\rs  = \nabla S_\rs, 
\eeq 
where the momentum potential $S$ originates from the classical HJ equation, %
and the potential $S_\rs$ - from the coordinate stochasticity. In the absence %
of stochasticity $S$ is the classical particle action that obey the HJ equation. %
We make the natural anzatz that the potential $S_\rs$ depends on $\bs{x}$ and $t$ %
via the coordinate probability distribution $\rho(\bs x,t)$ only: %
\, $S_\rs = S_\rs(\rho(\bs x,t))$.

Supposing that coordinate stochasticity induces new momentum part it is then %
natural to expect that the latter in turn will affect the particle action $S$. %
The simplest way to take into account this `feed back' is to suppose  that part %
of $S$ becomes $\rho$-dependent. We suppose that this part is proportional to %
$S_\rs$, and denote the difference by $S_-$. So that  we put 
\beq\lb{S_}
S = S_-  +   S_\rs(\rho),
\eeq 
and treat $S_-$ and $\rho$ as {\it independet fields}. Next we put $S = S_- + S_\rs(\rho)$ %
into the Reginatto action functional (\ref{Phi_B}) and apply the variational principle %
to the resulting functional 
\bea\lb{Phi_B'} 
\hspace{-5mm}\Phi_{B'} =  
\int \rho\,\left(\psl_t (S_- +   S_\rs) + \frac{1}{2m}\nabla (S_- +  S_\rs)\cdot 
\nabla(S_- +  S_\rs)  \right)\, \rd^3 x \,\rd t  \nonumber\\ 
\ftm \qquad \qquad  + \int \rho\left( V + 
\frac{\hbar^2}{8m}  \frac{1}{\rho^2} \nabla \rho\cdot \nabla \rho\right) \, 
\rd^3 x \,\rd t ,  
\eea
treating $\dlt S_-$ and $\dlt \rho$ as independent variations, vanishing at the end points. %
The resulting equations of the extremals read 
\beq\lb{eq S-1}
\frac{\psl \rho}{\psl t}+ \frac{1}{m}{\rm div} \left(\rho\,\nabla S_-  +   
\frac{\psl S_\rs }{\psl \rho}\nabla\rho \right)  = 0,
\eeq
\bea\lb{eq S-2}
\frac{ \psl S_-}{\psl t} + \frac{1}{2m}\left(\nabla (S_- +   S_\rs)\right)^2  - \frac{1}{m} 
\frac{\psl S_\rs}{\psl\rho} \left(  \nabla\rho\cdot\nabla S_- + \rho \nabla^2 S_-\right) \nonumber \\
 \ftm \qquad  - \frac{1}{m}  \frac{\psl S_\rs}{\psl\rho}\left(   \left(\rho \nabla^2\rho  + 
(\nabla\rho)^2\right) \frac{\psl S_\rs}{\psl\rho}  + 
\rho (\nabla\rho)^2 \frac{\psl^2 S_\rs}{\psl\rho^2}\right)  = 0.
\eea
Putting here $S_- = S -   S_\rs$ and using again the continuity equation we obtain %
the Bohm equations (\ref{Bohm eqs}) for $\rho$ and $S$, as desired.  Taking into account %
that $\sqrt{\rho}\exp(\ri S/\hbar)$ satisfies the Schr\"odinger equation we see that the %
$p_\rs$-potential $  S_\rs$ has the meaning of {\it $\rho$-dependent part} of the wave %
function phase.  

We note that this result is valid for any differential function $S_\rs(\rho)$ %
(but not if $S_\rs$ depends on $\psl \rho/\psl x$). %
One can use this freadom to subject $S_\rs$ to some desired constraints. %
One natural constraint is the vanishing average of $p_{\rs}$, $\ovl{p_\rs} = 0$. %
This can be satisfied with $S_\rs = \lam \ln (l^3\rho)$, where $l$ is a length parameter %
(so that $l^3\rho$ be dimensionless), i.e., 
\beq\lb{p_s}
 \bs{p}_\rs  =  \frac{\lam}{\rho}\nabla \rho . %
\eeq
This ensures the coincidence of total momentum average $\ovl{\bs{p}}$, %
$ \ovl{\bs{p}} := \int \rho(\bs{x},t)\bs{p}\,\rd^3x,$ %
with the average $\la \hat{p}\ra$ of  quantum momentum $\hat{p}$ in the corresponding %
state $\psi = \sqrt{\rho}\exp(\ri S/\hbar)$  (using the known  equality %
$\la\hat{p}\ra = \ovl{p_\rc}$ \cite{Martino94}). %
Furthemore we fix the parameter $\lam$ as $\hbar/2$, i.e. we use %
$\bs{p}_\rs =   \hbar(\nabla \rho)/2\rho$. 
To shorten the notation herefrom we consider the one-dimensional motion only. 

Formally the quantities $p_\rs/m$,  $\nabla S/m$ and $p = (p_\rs +\nabla S)/m$ coincide %
with the osmotic, current and forward velocities $u$,  $v$, $v_+$ in the Nelson SM, %
where many of their properties are thoroughly examined (see for example %
\cite{Martino93, Martino94, Martino95, Petroni99} and references there in).  %
Our interest here is focused on the properties related to the possibility $p_\rs$ to %
describe (at least partially) nonclassical fluctuations of the quantum-mechanical %
momentum $\hat{p}$. %
In this aim we compare statistical properties of $p_\rs$  with those of %
`nonclassical part' $\hat{p}_\nc$  of $\hat{p}$ \cite{Hall01},  \, %
$\hat{p}_\nc :=  \hat{p} - p_\rc$, $p_\rc = \psl S/\psl x$ . %
Hall \cite{Hall01} found the first two moments of $p_\nc$ as  %
\bea
 \la \hat{p}_\nc\ra=0, \qquad \la \hat{p}_\nc^2\ra= \la \hat{p}^2\ra - \la p_\rc^2\ra = 
(\Dlt \hat{p})^2 -  (\Dlt p_\rc)^2, \lb{p_nc means}\\
(\dlt x)^2 (\Dlt \hat p_\nc)^2 = \frac{\hbar^2}{4}, \lb{exact UR}  %
\eea  %
where $\dlt x$ is the Fisher length, $(\dlt x)^2 =  1/I_{\mbox{\tiny F}}$, the equality %
(\ref{exact UR}) being refered as `exact' UR. 
The above three properties are shared by $p_\rs$ too %
(established in terms of the osmotic momentum $mu$ earlier, see e.g. %
\cite{Falco82, Martino94} and references their in ): %
\bea
 \la p_\rs\ra=0, \qquad \la p_\rs^2\ra= \la \hat{p}^2\ra - \la p_\rc^2\ra  =  
(\Dlt \hat{p})^2 -  (\sig_{p_\rc})^2,\lb{p_s means}\\
(\dlt x)^2 (\sig_{p_s})^2 = \frac{\hbar^2}{4}, \lb{p_s exact UR} %
\eea %
where $\sig_{p_\rs}$, $\sig_{p_\rc}$ are variances of $p_\rs$, $p_\rc$: \, %
$(\sig_{p_\rc})^2 := \ovl{p_\rc^2} - {\ovl{p_\rc}}\,^2$. 
Further common properties of  $\hat p_\nc$ and $p_\rs$ could be revieled on examples of  %
some specific states only.   

We however have to note here an important difference in the properties of $\hat p_\nc$ and %
$p_\rs$: the linear correlation between $p_\rc$ and $p_\rs$ (i.e., the covariance $C_{p_\rs p_\rc}$ %
may not vanish, while the covariance $\Dlt_{p_\nc p_\rc}$ of $\hat p_\nc$ and $p_\rc$ vanishes %
in all states \cite{Hall01}. $C_{p_\rs p_\rc}$ may vanish in some specific families of states only, %
e.g.  in $\rho$ corresponding to the cannonical CS. (It is not vanishing  e.g. in squeezed states - %
see section 5).  
When $C_{p_\rs p_\rc} \neq 0$ the total second moment $p\equiv p_\rc +p_\rs$ is not %
equal to that of $\hat{p}$. With nonvanishing $p_\rs$-$p_\rc$ covariance the Hall and %
Reginatto scheme of derivation of ``quantum ensemble''  equations (i.e. Bohm equations %
(\ref{Bohm eqs})),  is not applicable. Therefore  {\it if ensemble interpretation} is %
applied to our scheme (with $p_\rs$) of derivation of Schr\"odinger equation, the %
resulting nonclassical ensemble could be called ``{\it semi-quantum}'' or, more briefly %
$p_\rs$-ensemble. And if one interpretes $p_\rc/m$ and $p_\rs/m$ as current and osmotic %
velocities respectively then Nelson SM scheme is applicable.

\section{R-S Type URs for Stochastic System}  

Inequalities of the type of  Robertson-Schr\"odinger UR (R-S URs) can be naturally 
and easily constructed for classical stochastic systems using the semi-definiteness of 
the covariance matrix (the matrix of dispersions \cite{Gned}) of two random quantities.
Gnedenko \cite{Gned} proved that all principal minors of the matrix of dispersions of 
any $n$ random quantities are nonnegative. 
For $n=2$ this means that the product of the two variances is greater or equal  to their 
squared covariance. 
Thus for any two random observables $\xi$, $\eta$ the following inequality is valid 
\beq\lb{ineq1} 
\sig_\xi^2\,\sig_\eta^2 \geq C^2_{\xi\eta}\, , 
\eeq 
where $\sig_{\xi}^2$ is the variances of $\xi$,  $\sig_{\xi}^2 = \ovl{\xi^2}-{\ovl{\xi}}\,^2$, 
and $C_{\xi\eta}$ is  the covariance, $C_{\xi\eta} = \ovl{\xi\eta}-\ovl \xi\,\ovl \eta$. 
Here $\ovl \xi$ is the mean value of $\xi$.
If the random quantity $\xi$ admits a probability density $\rho(\xi,t)$ one has 
$\ovl \xi = \int \rho(\xi,t) \xi\, \rd\xi$.
The inequality (\ref{ineq1}) is minimized iff $\xi$ and $\eta$ are linearly dependent \cite{Gned}. 
For brevity the stochastic quantity and its values are denoted with the same letter.

We see that the inequality (\ref{ineq1}) is of almost the same form as the R-S UR 
(\ref{SUR}) in quantum mechanics, the mean commutator of the two observables 
being missing only. 
Therefore the inequalities of the form (\ref{ineq1}) in stochastic mechanics and in any %
probability theory could be naturally called the R-S type URs. 
For given two quantities $\xi$, $\eta$ such inequality should briefly be referred to as $\xi$-$\eta$ UR.

Next we construct and discuss the R-S type URs for the coordinate and momentums of the stochastic particle. 
In the 'semi-quantum ensemble' interpretation we have to treat $p_\rs + p_\rc \equiv p$ as total %
particle momentum and compare the $x$-$p$ UR with $\hat x$-$\hat{p}$ UR in quantum mechanics. %
Similarly URs between any other pair of  the set $(x,\,p_\rc,\, p_\rs,\, p)$ is to be compared with 
UR of  the corresponding quantum pair from $(\hat x,\,\hat p_\rc,\, \hat{p}_\nc,\, \hat{p})$. 
In the stochastic mechanics interpretation   the set $(x,\,p_\rc,\, p_\rs, \, p)$ coinsides with %
$(x,\,mv,\, mu,\, mv_+)$, where $v,\,u,\,v_+$ are current, osmotic and forward velocities. 

In stochastic mechanics $x$-$mu$ UR (the osmotic UR) was established in \cite{Cetto72} and 
\cite{Falco82} in the `Heisenberg form'  $(\sig_x)^2 (\sig_{mu})^2 \geq \hbar^2/4$, which we  
rewrite as 
\beq\lb{os UR} %
(\sig_x)^2 (\sig_{p_\rs})^2 \geq \hbar^2/4.
\eeq %
In \cite{Martino93, Martino94} the osmotic inequality  was extended to the processes with non %
constant diffusion coefficient $\nu(x,t)$ in the form   $(\sig_x)^2 (\sig_{u})^2 \geq \ovl{\nu}\,^2$. %
Comparing (\ref{os UR}) with  (\ref{ineq1}) we see that the squared $x$-$p_\rs$ covariance is %
universally constant and equaled to $\hbar^2/4$. The covariance itself is 
\beq\lb{C_xps} 
C_{xp_{\rs}} =  -\hbar/2.
 \eeq
In Heisenberg UR in quantum mechanics the universal term $\hbar^2/4$ comes from 
the nonvanishing commutator of coordinate and momentum operators. 
We now see that in SM and in `quantum ensemble' aproach  this term
comes from the $x$-$p_{\rs}$ covariance.  
The constancy of the covariance $C_{x p_{\rs}}$ is, in fact, due to the vanishing
first moment  of our $p_{\rs}$. 
Due to this property the variance of $p_{\rs}$ is proportional to the Fisher information 
(as required in \cite{Hall02} for the variance of their `nonanalyzable' $\bs p_\rN$), and 
 the $x$-$p_{\rs}$ UR (\ref{os UR}) coincides with the known Cramer-Rao inequality  
$\sig_x^2\,I_{\rm F} \geq 1$. 
For Gaussian $\rho(x,t)$ one has $I_{\rm F}(\rho) = 1/\sig_x^2$ \cite{NF}. 
Therefore for Gauss distribution the UR (\ref{os UR}) is minimized along with the 
Cramer-Rao inequality.
The UR (\ref{os UR}) is to be compared with the $\hat x$-$\hat{p}_\nc$ UR  
$(\Dlt \hat x)^2 (\Dlt \hat{p}_\nc)^2 \geq \hbar^2/4$  \cite{Hall01} and with the chain 
inequalities (\ref{HUR0}).

Unlike $C_{xp_{\rs}}$ the covariances of other pairs of  the  set 
$\{x, p_\rc, p_{\rs}, p\}$, though having to obey the R-S type URs 
(\ref{ineq1}), do not take universally fixed values. 
In the next section we shall discuss this on the (one-dimensional) examples, %
comparing the calculated moments with the corresponding ones in quantum mechanics.

\section{Examples: Coherent States and Squeezed States}  

In these section we calculate the first and second moments of $x$, $p_{\rs}$, $p_\rc$ 
and $p=p_{\rs}+p_\rc$ and  the related R-S types of URs in `stochastic states' 
$\rho(x,t)$ corresponding to the Glauber CS and  canonical SS  in quantum mechanics, 
and compare them with the related quantum moments. 
Nelson stochastic mechanics (SM) images of CS and SS have been discussed previously 
in several  papers: of CS in \cite{Guerra81} - \cite{Martino95}, %
and of  SS and CS - in \cite{Petroni99}-\cite{Martino95} in the context of  `stochastic mechanics 
and control theory'.  
Here we write these images and the related moments and URs in more standard quantum 
optical and quantum mechanical parameters (see e.g.  \cite{ T01}-\cite{Dod02} \cite{Yuen}).  
\medskip

\begin{itemize}

\item a) {\bf Glauber coherent states.} Glauber CS \cite{G} are defined
as eigenstates $|\alf\ra$ of the boson annihilation operator $\hat{a}$,
\beq\label{a}
\hat{a}|\alf\ra = \alf|\alf\ra, \qquad \hat{a} =
\frac{1}{\sqrt{2\hbar}}\left(\hat{x}\sqrt{m\ome}+
\frac{\ri}{\sqrt{m\ome}}\hat{p}\right)
\eeq
where $\hat x$ and $\hat p$ are coordinate and momentum
operators, and $m$ and $\ome$ are parameters of dimension of mass
and frequency correspondingly. For the harmonic oscillator $m$ is the
mass of the particle, and $\ome$ is the oscillator frequency.
These CS have been introduced by Glauber in 1963 \cite{G} and are
known as the {\it most classical} quantum states.  In $|\alf\ra$ the first
and second moments of $\hat{x}$ and $\hat{p}$ read
\beq\label{in CS 1}
\la\alf|\hat{x}|\alf\ra =
\sqrt{\frac{2\hbar}{m\ome}}\, {\rm Re}\,\alf,\qquad
\la\alf|\hat{p}|\alf\ra =  \sqrt{2\hbar m\ome}\, {\rm Im}\alf
\eeq
\beq\label{in CS 2}
(\Dlt \hat{x})^2  = \frac{l^2}{2},\qquad (\Dlt\hat{p })^2 = \frac{\hbar^2}{2l^2},
 \qquad \Dlt_{xp}  = 0. 
\eeq
where $l^2 = \hbar/m\ome$ (the length parameter).  We see that the moments  minimize 
R-S UR (\ref{xp SUR}) on the lowest possible level (which is the equality in the Heisenberg UR): 
$(\Dlt\hat{x})^2 (\Dlt\hat{p})^2  =\hbar/4$. 


To perform the comparison with the moments in `semiclassical ensemble' and in SM 
we need the time-dependent CS, i.e., eigenstates of $\hat{a}$ that
obey the  Schr\"odinger equation. The first
requirement can be met if the CS wave function depends, up to a
$x$-independent phase factor, on $t$ through the eigenvalue
$\alf$: $\psi_\alf(x,t) = \exp(\ri\vphi(t))\psi_{\alf(t)}(x)$, $\hat{a}\psi_\alf(x,t) = 
\alf(t)\psi_\alf(x,t)$. Such
{\it stable} CS $\psi_{\alf}(x,t)$ exist for the stationary harmonic oscillator Hamiltonian,
$\hat{H} = -(\hbar^2/2m) \psl_{xx}+(m\ome^2/2) x^2$ 
with $\alf(t)=\alf\exp(-\ri \ome t)$,  $\vphi(t) = -\ome t/2$ and  
\beq\label{psi_cs}
\hspace*{-10mm} \psi_{\alf(t)}(x) = \left( \frac{1}{\pi l^2}\right)^{\frac 14}
\exp\left[-\frac{1}{2}\left(\frac{x}{l} -  \alf(t) \sqrt{2}\right)^2 +
\frac{1}{2} \left(\alf^2(t) - |\alf(t)|^2\right)\rt], 
\eeq
where $l = \sqrt{\hbar/m\ome}$\, (the length parameter). For the stable CS
$\psi_{\alf(t)}(x)$ the first and the second moments of $\hat{x}$
and $\hat{p}$ are given by the same formulas (\ref{in CS 1}),
(\ref{in CS 2}) but with the time-dependent eigenvalue $\alf(t)$.

Next we put $|\psi_{\alf(t)}(x)|^2 = \rho_{cs}(x,t)$ and calculate
the stochastic moments of $x$ and $p_{\rs}$.  
Formula in (\ref{psi_cs}) readily shows that 
(we put $\alf_1={\rm Re}\,\alf,\,\,\, \alf_2={\rm Im}\,\alf $)  
\beq\lb{ovl x} %
\ovl x = l\sqrt{2}\alf_2(t)\quad {\rm and}\quad \sig_x^2 = \frac{l^2}{2},%
\eeq
and provides $p_{\rs} = -\hbar\left(x - \ovl x\right)/l^2$, wherefrom 
\beq\lb{ovl ps} %
 \ovl p_{\rs} = 0,\quad {\rm and}\quad \sig_{p_{\rs}}^2 = \frac{\hbar^2}{2l^2} \cdot  %
\eeq %
We see that $x$-$p_{\rs}$ UR (\ref{os UR}) is minimized \cite{Martino93}: 
$\sig_{x}^2 \sig_{p_{\rs}}^2 = \hbar^2/4$.

To find the first and second moments of $p_\rc$ and the `total
momentum' $p = p_\rc + p_{\rs}$ we need the action $S(x,t)$, %
\beq\label{S} %
S(x,t) = \frac{\hbar}{l} \alf_2(t)\sqrt{2}\,x - \hbar \alf_1(t)\alf_2(t) -
\hbar\frac{\ome t}{2}\cdot %
 \eeq %
 Then we get $p_\rc = \hbar \sqrt{2}\alf_2(t)$ and %
\beq\label{mean p_c} %
 \ovl p = \overline{p_\rc} = \int p_\rc\,
\rho_{cs}(x,t) \, \rd x = \frac{\hbar}{l}\sqrt{2}\alf_2(t), %
\eeq %
verifying the known coincidence of $\ovl x$ and $\ovl p$ with quantum
means $\la\hat{x}\ra$ and $\la \hat{p}\ra$ \cite{Falco82,Martino93}. 

Next we calculate the second moments of $p_\rc$ and $p$ and the
related covariances. The covariance $C_{x p_{\rs}}$, as noted in the
previous section, is universally equal to $-\hbar/2$. The correlation 
between $p_\rc$ and $p_{\rs}$ in $\rho_\rcs$ turned out to be vanishing: 
\beq\label{C_pspc} %
C_{p_\rc p_{\rs}} = \ovl{p_\rc p_{\rs}} = 0.%
\eeq %
Thus the required in \cite{Hall02} properties (\ref{N}) of the `nonanalyzable' 
momentum $\bs p_\rN$ are satisfied by $p_{\rs}$ in $\rho_\rcs$ and the first two 
moments of  $p_\rs$ and $\hat{p}_\nc$ and that of $p$ and $\hat{p}$ do coincide.
The third and higher moments of $p_\rs$ and $\hat{p}_\nc$, however are found to 
coinside at all times in the ground state only. For example $\la p_\rs^3\ra = 0$, 
while $\la \hat{p}_\nc^3\ra = \la \hat p\ra^3 + \la\hat p\ra/2- \la\hat p\ra^2 $, %
$\la\hat p\ra =  \sqrt{2}{\rm Im}\alf(t)$.

For the rest two variances and covariances in $\rho_\rcs$ we get
\beq\label{Cxp rhocs} %
\bt{l}
$\dstyle C_{xp_\rc} = 0,\qquad C_{xp} = C_{xp_\rc} + C_{xp_{\rs}}= -\hbar/2$,\\[2mm]
$\dstyle \sig_{p_\rc} =0, \qquad \sig_p^2 = \sig_{p_{\rs}}^2 = \frac{\hbar^2}{2l^2}\cdot$ %
\et
\eeq
Note the vanishing variance of the momentum $p_\rc:= \psl S/\psl x$ in $\rho_\rcs$.
As we shall see below this is again a particular property of $\rho_{\rcs}$.

Now one can easily check that the R-S type URs (\ref{ineq0}) for
all the coordinate-momentum pairs $x$-$p_\rc$, $x$-$p_{\rs}$ and
$x$-$p$ are minimized in $\rho_\rcs$. In particular the chain inequalities 
\cite{Cetto72,Falco82} %
\beq\lb{chain UR}
(\Dlt \hat x)^2  (\Dlt  \hat{p})^2 \geq \sig^2_ x \sig^2_{p_\rs} \geq \hbar^2/4
\eeq
are also minimized. 
These minimizations follow from the fact that in $\rho_{\rcs}$ all quantities %
$x,\, p_\rs,\, p_\rc$ are linearly dependent \cite{Gned}. We note that in %
terms of the SM velocities $u=p_\rs/m$, $v= {\rm grad} S/m$ all the above 
CS-related moments and URs (\ref{os UR}), (\ref{chain UR}) were considered %
previously \cite{Cetto72,Falco82} \cite{Martino93, Martino95}.       
\medskip

In quantum mechanics CS $|\alf\ra$ are regarded as the `most
classical' states.  They can be uniquely determined as states minimizing 
the inequality  \cite{T01} %
\beq\lb{hur2} %
(\Dlt \tilde{\hat x})^2 + (\Dlt \tilde{\hat{p}})^2 \geq 1, %
\eeq %
where $\tilde x$ and $\tilde{\hat{p}}$ are dimensionless coordinate and momentum.
One can see that in $\rho_\rcs$ the sum $\sig^2_{\tilde x}+\sig^2_{\tilde p}$ %
 also equals unity. However in other states the inequality %
$\sig^2_{\tilde x}+\sig^2_{\tilde p}\geq 1$ %
may be violated, as we shall see on the example of squeezed states. \\

\item b) {\bf Squeezed States}. Squeezed states (SS) are defined as
quantum states in which the variance (uncertainty) of coordinate or
the variance of the momentum is less than its value in the ground
state of the oscillator. The SS are known as {\it nonclassical states}
since they exhibit many nonclassical properties. The famous example of  SS
are the eigenstates of the linear combination of Bose creation and
annihilation operators $\tld u\hat{a} + \tld v\hat{a}^\dg \equiv \hat A$ \cite{Yuen}, %
which we rewrite in terms of $\hat{x}$ and $\hat{p}$ as $\mu \hat{x}/l +
 \ri\nu l\hat{p}/\hbar$ $(\mu = (\tld u+\tld v)/\sqrt{2},\,\, \nu =(\tld u-\tld v)/\sqrt{2}$), %
\beq\label{SS} 
(\mu \hat{x}/l + \ri\nu l\hat{p}/\hbar)|\alf;\mu,\nu\ra = \alf|\alf;\mu,\nu\ra, %
\eeq
where $\alf$ is a complex number,  $l$ is the length parameter, and %
\beq\lb{mu nu}     
|u|^2-|v|^2 = 2{\rm Re}(\mu^*\nu) = 1,
\eeq
which are to ensure $[\hat A,\hat A^\dg]=1$.  It was noted in \cite{T94} that SS
$|\alf;\mu,\nu\ra$ are states that minimize the R-S UR and
coincide with the `correlated CS' of ref. \cite{DKM}. That is why
they are also called generalized intelligent states or R-S
intelligent states \cite{T94}.

In the coordinate representation the SS wave functions take the
form of exponential of a quadratic. These states are time-stable
for any quadratic in $\hat x$ and $\hat p$ Hamiltonian, in particular for %
the harmonic oscillator with constant or time-dependent
frequency $\ome(t)$. The normalized time-dependent
wave function of an initial SS $|\alf;\mu_0,\nu_0\ra$ reads
\begin{eqnarray}\label{psi_ss}     
 \psi_{\alf \mu_0 \nu_0}(x,t)
 &=&\left(l\nu(t)\sqrt{2\pi}\right)^{-\frac{1}{2}}
\exp\left[-\frac{\mu(t)}{2 l^2\nu(t)}\left(x - \frac{ l}{\mu(t)}\alf
\right)^2\right]\nonumber\\
&&\\
&& \times \exp\left[ - \frac{1}{2}\left(|\alf|^2 - \frac{\mu^*(t)}{\mu(t)}\alf^2
\right)\right] \nonumber
\end{eqnarray}
where $\alf$ is constant, and $\mu(t)$, $\nu(t)$ ($\mu(0)=\mu_0,\, \nu(0)=\nu_0$) 
satisfy certain first order equations, which can be reduced to the
classical harmonic oscillator equation $\ddot{\eps} +
\ome(t)^2\eps =0$ through the substitutions $\mu(t) =
-\ri\dot{\eps}/\sqrt{2\ome_0}$, $\nu(t) = \eps \sqrt{\ome_0/2}$,
$\ome_0$ being constant of  inverse time dimension \cite{DM89, T01}. %
With such $\mu(t),\, \nu(t)$ the operator ${\small\hat A}$ is a dynamical invariant 
of the nonstationary oscillator, i.e., {\small $d \hat A/dt = 0$}.   
The family of stable SS includes the family of CS as a subset:  %
If $\eps(0) = 1/\sqrt{\ome_0}$ and $\dot{\eps}(0) = \ri\sqrt{\ome_0}$
(that is $\mu_0\equiv \mu(0)=1/\sqrt{2},\, \nu_0\equiv
\nu(0)=1/\sqrt{2}$) then the wave function (\ref{psi_ss})
represents the time-evolution of an initial Glauber CS $|\alf\ra$.
In fact, in terms of $\eps$, $\dot{\eps}$ the wave functions
(\ref{psi_ss}) have been constructed and discussed earlier in
\cite{ MMT} as time evolved CS for quadratic systems.

The first and the second moments of $\hat{x}$ and $\hat{p}$ in SS
(\ref{psi_ss}) read \cite{DM89, T01}
\beq\label{SS m1}     
 \la \hat{x}\ra =  2l{\rm Re}(\alf(t)\nu^*(t)),\qquad
\la \hat{p}\ra = 2\frac{\hbar}{l}{\rm Im}(\alf(t)\mu^*(t)) .
\eeq
 \beq \label{SS m2}   
 \hspace*{-14mm}
(\Dlt \hat x)^2 = l^2|\nu(t)|^2,\quad \,\,
 (\Dlt \hat p)^2 = \frac{\hbar^2}{l^2}|\mu(t)|^2, \quad\,\,
 \Dlt_{xp} = \hbar {\rm Im}(\mu^*(t) \nu(t))
\eeq
the  second moments saturating the R-S UR (\ref{xp SUR}).

To calculate the stochastic moments in $\rho_{ss}= |\psi_{\alf
\mu_0 \nu_0}(x,t)|^2$ we have to find the momentum potentials $S$
and $S_\rs = (\hbar/2) \ln\rho_{ss}$ (furthermore we skip the
argument $t$ of $\alf(t)$, $\mu(t)$ and $\nu(t)$):

\beq\lb{S_ss}    
S(x,t) =  -\frac{\hbar}{2l^2}{\rm
Im}\left(\frac{\mu}{\nu}\right)x^2 +
 \frac{\hbar}{l} {\rm Im}\left(\frac{\alf}{\nu}\right)x  + g_1(t), 
\eeq

\beq\lb{rho_ss}    
\rho_{ss}(x,t) =
\frac{1}{|\nu|l\sqrt{2\pi}}\exp\left[-\frac{1}{2l^2|\nu|^2}
\left[x - 2l {\rm Re}(\alf\nu^*)\right]^2 + g_2(t)\right] ,
\eeq
where the terms $g_1(t),\,g_2(t)$ are $x$-independent,
\[ g_1(t) =  - \frac{1}{2}{\rm Im}\left(\frac{\alf^2}{\mu\nu}\right) -
\frac{1}{2}{\rm arg}(\nu) +\frac{1}{2}{\rm
Im}\left(\frac{\alf^2\mu^*}{\mu} \right), \]
\[ g_2(t)  =  \frac{2}{|\nu|^{2}} {\rm Re}^2(\alf\nu^*) + {\rm Re}\left(\alf^2/\mu\nu\right) 
+ {\rm Re}\left(\mu^*\alf^2/\mu\right) - |\alf|^2 \, .\]

The first moments of $x$ and $p$ in $\rho_{ss}$ coincide with the
quantum means $\la \hat{x} \ra$, $\la \hat{p}\ra$, eqs.(\ref{SS m1}).
For the second moments of $x,\,p_\rc,\,p_\rs,\, p=p_\rs+p_\rc$ we find
\beq\label{SS Cov's}    
\bt{l}
$\dstyle C_{x p_\rc} = -\hbar {\rm Im}(\mu\nu^*), \qquad  
C_{p_{\rs} p_\rc} = \frac{\hbar^2}{2l^2|\nu|^2} {\rm Im}(\mu\nu^*)$,\\[1mm]
$\dstyle C_{x p} =- \hbar\left(\frac{1}{2} +{\rm Im}(\mu\nu^*)\right)$,   
\et
\eeq

\beq\label{rhoSS m2}      
\bt{l} 
$\dstyle \sig_x^2 = l^2|\nu|^2, \qquad \sig_p^2 =
\frac{\hbar^2}{l^2|\nu|^2}\left(\frac{1}{2} + {\rm
Im}(\mu\nu^*)\right)^2$\\  $\dstyle \sig_{p_\rc}^2 =
\frac{\hbar^2}{l^2} |\mu|^2 \sin^2(\dlt\vphi),
\qquad \sig_{p_{\rs}}^2 = \frac{\hbar^2}{4l^2|\nu|^2}$,
\et
\eeq
 where $\dlt \vphi = \arg \mu - \arg \nu$. From
(\ref{SS Cov's}) and (\ref{rhoSS m2}) it follows that the
R-S URs for {\it all pairs} of observables  $x$, $p_c$, $p_{\rs}$,
$p$  {\it are minimized} in $\rho_{ss}$,
\beq\label{all URs1}
\hspace{-5mm}\sig_x^2 \sig_{p_\rc}^2  =  C_{xp_\rc}^2 = \hbar^2 {\rm
Im}^2(\mu\nu^*), 
\quad \sig_x^2 \sig_{p}^2  =  C_{xp}^2 =
\hbar^2 \left(\frac{1}{2} + {\rm Im}(\mu\nu^*)\right)^2 ,
\eeq
\beq\lb{all URs2}
 \sig_x^2 \sig_{p_{\rs}}^2  = C_{xp_{\rs}}^2 = \frac{\hbar^2}{4}, 
\qquad \sig_{p_c}^2 \sig_{p_{\rs}}^2  =  C_{p_cp_{\rs}}^2
= \frac{\hbar^4}{4l^4|\nu|^4} {\rm Im}^2(\mu\nu^*), 
\eeq
as expected due to the linearity of  $p_c, p_{\rs},p$ in terms of $x$. 

In $\rho_{ss}$ however, unlike the case of $\rho_{cs}$,  the
dimensionless variances of $x$ and momentum  $p$ (or $p_{\rs}$) are
no more equal and none of the stochastic momentum uncertainties
coincides identically with the quantum uncertainty $\Dlt \hat p$.
These second moment's differences could be 
interpreted as due to the {\it `nonclassicality'} of the SS.   
The calculations shows that the variance of `semi-quantum ensembe' moment 
$p = p_\rs + p_\rc$  can be greater or less than $(\Dlt\hat p)^2$. The ratio 
$$ r_p = \left[(\Dlt\hat p)^2 - \sig_p^2\right] /(\Dlt\hat p)^2$$ %
could be used to described the deviation of momentum fluctuations in 
$p_\rs$-ensemble state $\rho_\psi$ from quantum fluctuations in $\psi$. 
For SS $\psi_{\alf \mu \nu}$ it takes the form,   
\beq\lb{r_p}
 r_p = -\frac{{\rm Im}(\mu\nu^*)}{|\mu\nu|^2}, 
\eeq 
its value oscillating between $\pm 1$.  It shows that the two variances coinside 
in states with $\mu$ and $\nu$ phase difference equal to $2n\pi$.  
Due to the nonvanishing covariance $C_{p_\rc\, p_\rs}$ the variance of $p$ 
may vanish for certain values of $\mu,\, \nu$. In such states the variances 
$\sig_x^2$ and $\sig_p^2$ could not preserve the inequalities (\ref{HUR}) 
and (\ref{hur2}).   

\end{itemize}

\medskip

\section*{Conclusion}

It has been shown that the Bohm equations, which are equivalent to the Schr\"odinger equations, %
can be derived from the classical Hamilton-Jacobi equation admiting an additional particle %
momentum $p_\rs$ of the form of stochastic mechanics osmotic momentum and using the %
variational principle with appropriate change of variables. The variational functional is similar %
to that of Reginatto and Hall \cite{Regi98, Hall01} which incorporates Frieden \cite{Frieden} %
principle of minimal Fisher information.  

The fluctuations of  $p_\rs$, classical momentum  
$\psl S/\psl x \equiv p_\rc$ and the `total particle momentum' $p= p_\rs+p_\rc$ and the %
related Robertson-Schr\"odinger type uncertainty relations (URs) are examined and compared %
with the corresponding quantum ones on the example of canonical coherent and squeezed states  %
(CS and SS). In CS the uncertainties (the variances) of  $p$ and quantum $\hat p$ and the related %
URs coincide, while in SS they reveal differences. The latter are due to the nonvanishing %
$p_\rs$-$p_\rc$ correlations. The normalized deviation of  variance of $p$ from that of $\hat p$ %
however is bounded between $\pm1$. Thus in the ensemble interpretaion, our `$p_\rs$-ensemble' 
can only approximately and partially reproduce statistical properties of  `quantum ensemble'.  %
The correspondence with the Nelson stochastic mechanics is obtained via the identification   
of $p_\rs/m$,  $p_\rc/m$ and $p/m$  with the osmotic, current and forward particle velocities.

\end{document}